\documentclass[aps,onecolumn,showpacs]{revtex4}

\usepackage{graphicx}
\include{amssym}
\def\PSfig#1#2{\centerline{\scalebox{#1}{\includegraphics{#2}}}}

\begin{document}

\title{Explicit chiral symmetry breaking and the $1/N$ expansion in 
       models with multi-fermion interactions} 

\author{A. A. Osipov}
\affiliation{Dzhelepov Laboratory of Nuclear Problems, JINR, 
         141980 Dubna, Russia}
\author{B. Hiller, A. H. Blin}
\affiliation{Centro de F\'{\i}sica Te\'{o}rica, Departamento de F\'{\i}sica
        da Universidade de Coimbra, 3004-516 Coimbra, Portugal}
\date{\today}

\begin{abstract}
Explicit chiral symmetry breaking is a natural feature of many
QCD inspired models with multi-quark interactions. To carry out the 
$1/N$ expansion of these theories, one integrates over the quark 
f\mbox{}ields. In the process of calculating the one-loop fermion 
determinant which is not chiral invariant due to quark masses, special 
care must be taken in order not to alter the original symmetry
breaking pattern of the Lagrangian. We show how to do this
consistently.
\end{abstract}

\pacs{11.30.Rd, 12.39.Fe}

\maketitle



\section{Introduction}

The purpose of this paper is to resolve a problem that arises in 
models with chirally symmetric multi-fermion interactions when this
symmetry is explicitly broken by the mass term of the fundamental
fermion f\mbox{}ield. These can be, in particular, the corresponding 
extensions of the $SU(N)$ Thirring model in two space-time dimensions 
\cite{Gross:1974}, or the Nambu -- Jona-Lasinio like models in four 
space-time dimensions \cite{Nambu:1961,Bijnens:1993}. To carry out the 
$1/N$ expansion of these theories, one should integrate over the 
fermions, facing the task of calculating the one-loop fermion diagrams
to obtain the ef\mbox{}fective action of the auxiliary bosonic 
f\mbox{}ields. The problem is that the standard procedure leads to a 
result which does not possess the transformation properties of the 
original fermion Lagrangian with respect to the action of the 
continuous chiral group. 

The essence of the problem has been already discussed in Ref. 
\cite{Osipov:2000}, where the formal solution has been found by adding
a functional $P$ in the bosonic f\mbox{}ields and their derivatives to
the real part of the fermion determinant. Here we suggest an
alternative method to solve the problem. This new solution leads
directly to the result and in addition provides for an interpretation
of the correcting functional $P$ in terms of Feynman diagrams.


\section{The $1/N$ expansion of the $SU(N)$ Thirring model with explicitly
         broken chiral symmetry}

The $SU(N)$ Thirring model describes a system of $N$ Dirac 
f\mbox{}ields $\psi_{(a)}$ carrying the flavor index $a=1,2,\ldots N$,
which we promptly suppress, in two space-time dimensions with the
Lagrangian density 
\begin{equation}
\label{GN1}
   {\cal L} (\psi, \bar{\psi}) =   
             -\bar{\psi}(i\!\!\not\!\partial
             +\widehat{m})\psi + \frac{g^2}{2}
             \left[(\bar\psi\psi )^2+(\bar\psi i\gamma_5\psi )^2
             \right], 
\end{equation}
where $\bar\psi=\psi^\dagger\beta$, $\not\!\!\partial =\gamma^\mu
\partial_\mu$. The $2\times 2$ antihermitian symmetric matricies 
$\alpha^\mu =-i\beta\gamma^\mu$, $\mu =0,1$ generate the 
Clif\mbox{}ford algebra: $\{\alpha^\mu , \alpha^\nu\}=2
\delta_{\mu\nu}$. The matrix $\beta$ is imaginary and antisymmetric. 
We take here the choice $\beta =-\gamma^0=\sigma_2$,
$\gamma^1=i\sigma_1$, $\gamma_5=\sigma_3$, where $\sigma_i$ are the
Pauli matricies. One has also $\{\gamma^\mu , \gamma^\nu\}=
2g^{\mu\nu}$ with the Minkowski metric tensor $g^{\mu\nu} = 
\mbox{diag}(1,-1)$.
 
The Lagrangian possesses a $U(1)$ chiral symmetry $\psi_{(a)}\to 
e^{i\theta\gamma_5}\psi_{(a)}$ which is explicitly broken by the mass
term. One can easily f\mbox{}ind that under small variations 
$\theta\ll 1$, we have
\begin{equation}
\label{dL1}
   \delta {\cal L} (\psi, \bar{\psi}) =
   -\widehat{m}\delta (\bar\psi\psi )
   = -2\theta\widehat{m}\bar\psi i\gamma_5\psi .
\end{equation} 

Following Ref. \cite{Gross:1974}, one can equivalently introduce
auxiliary f\mbox{}ields $\sigma$, $\phi$, and write
\begin{equation}
\label{GN2}
   {\cal L} (\psi, \bar{\psi}, \sigma, \phi ) =  
            -\bar{\psi}(i\!\!\not\!\partial
            +\widehat{m} +\sigma +i\gamma_5\phi )\psi 
            -\frac{\sigma^2+\phi^2}{2g^2}\ . 
\end{equation}
The Euler -- Lagrange equations for the bosonic f\mbox{}ields are
constraints
\begin{equation}
   -\frac{\partial {\cal L}}{\partial\sigma}=\bar\psi\psi
   +\frac{\sigma}{g^2}=0, \qquad
   -\frac{\partial {\cal L}}{\partial\phi}=\bar\psi i\gamma_5\psi
   +\frac{\phi}{g^2}=0,  
\label{sfe} 
\end{equation}
which relate the chiral transformations of fermions and bosons, and 
hence 
\begin{eqnarray}  
\label{dsigma}
   \delta\sigma &=& -g^2\delta (\bar\psi\psi ) = 2\theta\phi , \\
   \delta\phi   &=& -g^2\delta (\bar\psi i\gamma_5\psi ) 
                 =  -2\theta\sigma .
\label{dphi}
\end{eqnarray}
Then it follows from (\ref{dL1}) and (\ref{dsigma}) that the
inf\mbox{}initesimal symmetry transformation is now
\begin{equation}
\label{dL2}
   \delta {\cal L} 
   =\frac{\widehat{m}}{g^2}\,\delta\sigma .
\end{equation} 

Performing the Gaussian integral over the fermions in the functional 
integral corresponding to (\ref{GN2}) (see Eq. (\ref{Z1}) below), 
one f\mbox{}inds that the real part of the ef\mbox{}fective action 
$S_{{\rm eff}}$ has a dif\mbox{}ferent transformation property. 
Indeed, one obtains in euclidean space
\begin{equation}
\label{RS}
   \mbox{Re}\,S^{{\rm eff}}_E = \frac{N}{2} \mbox{Tr}\ln 
   D^\dagger_ED_E -\!\int\!{\rm d}^2x_E\,
                \frac{\sigma^2+\phi^2}{2g^2}
\end{equation} 
with
\begin{equation}
\label{dS1}
   \delta\left(\mbox{Re} S^{{\rm eff}}_E\right)= \widehat{m}N\,
   \mbox{Tr} \left(\frac{2\theta\phi}{D^\dagger_ED_E}\right),
\end{equation} 
where the Dirac operator is given by 
$   D_E = -(i\!\!\not\!\partial +\widehat{m} +\sigma +i\gamma_5
         \phi ). 
$
Here we use the following convention: a Lorentz 2-vector $x^\mu$ is
continued as $x^0\to -ix^E_2$, $x^1\to x^E_1$. The euclidean
$\gamma$-matrices $\gamma^E_a$ are antihermitian $\gamma^0\to
\gamma_2^E=-i\sigma_2$, $\gamma^1\to\gamma^E_1=i\sigma_1$. They 
satisfy the anticommutation relation $\{\gamma_a^E,\gamma_b^E\} = -2
\delta_{ab}$. Accordingly $\not\!\partial =\gamma_a^E\partial^E_a$ is 
a hermitian operator. Furthemore $\mbox{Tr}=\!\int\!{\rm d}^2x_E
\mbox{tr}$, where the last trace is over Dirac indices.     

The unsuccessful outcome of Eq. (\ref{dS1}) shows that the standard 
evaluation of the Gaussian functional integral over fermions, when
chiral symmetry is explicitly broken, must be corrected, otherwise the 
result is not consistent with the constraint imposed by the symmetry 
breaking pattern of the theory at the tree level. In particular, this
can be done by the insertion of new counterterms into the 
ef\mbox{}fective action (\ref{RS}). The corresponding method has been 
developed in \cite{Osipov:2000}. However, it was not clear then (a) if 
these counterterms were compatible with the dynamics of the original
multi-fermion system, and (b) if the method could be applied to  
renormalizable models. In the following we shall f\mbox{}ind the 
answers to these questions describing a new way of dealing with the 
problem.


\section{Tadpole mechanism}
\label{sec.tm}
\begin{figure}[b]
   \PSfig{0.7}{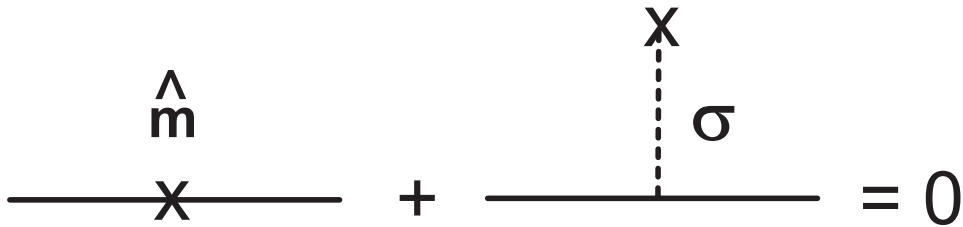}
   \caption{\small The mass term of the fermion is cancelled due to a 
   contribution generated by the $\sigma\to\mbox{vacuum}$ transition.} 
\label{fig1}
\end{figure}

It is convenient to divide the Lagrangian (\ref{GN2}) into two parts 
${\cal L}={\cal L}_1+{\cal L}_2$. The f\mbox{}irst part 
\begin{equation}
\label{GN3}
   {\cal L}_1 = -\bar{\psi}(i\!\!\not\!\partial
            +\widehat{m} +\sigma +i\gamma_5\phi )\psi 
            -\frac{\widehat{m}\sigma}{g^2} 
\end{equation}
is invariant under the action of the chiral group. We expect that this 
property will be still fulf\mbox{}illed for the corresponding part of
the ef\mbox{}fective Lagrangian obtained as a result of integration
over the $\psi$ f\mbox{}ields. Our expectation is based on the
observation that the dangerous symmetry breaking fermion mass term can 
be easily subtracted if one considers the tree-level tadpole $\sigma\to
\mbox{vacuum}$ transition contained in ${\cal  L}_1$, as it is shown 
in Fig. \ref{fig1}.  

The second part 
\begin{equation}
\label{GN4}
   {\cal L}_2 = \frac{\widehat{m}\sigma}{g^2}
                -\frac{\sigma^2+\phi^2}{2g^2} 
\end{equation}
possesses the necessary transformation property $\delta {\cal L}_2 = 
\widehat{m} \delta\sigma /g^2$ of the Lagrangian (\ref{GN2}). Note,
that the term $\widehat{m}\sigma /g^2$, which we subtract in
(\ref{GN3}) and add in (\ref{GN4}), is unambiguously determined by the 
chiral symmetry restriction $\delta {\cal L}_1=0$. It follows from  
${\cal L}_2$ additionally that the auxiliary f\mbox{}ields $\sigma$
and $\phi$ do not propagate at tree level. They have the $\delta$-like 
Green functions   
\begin{equation}  
\label{prop}
   \Delta_{\sigma ,\phi}(x-y)=-ig^2\delta (x-y).
\end{equation}

\begin{figure}[b]
   \PSfig{0.7}{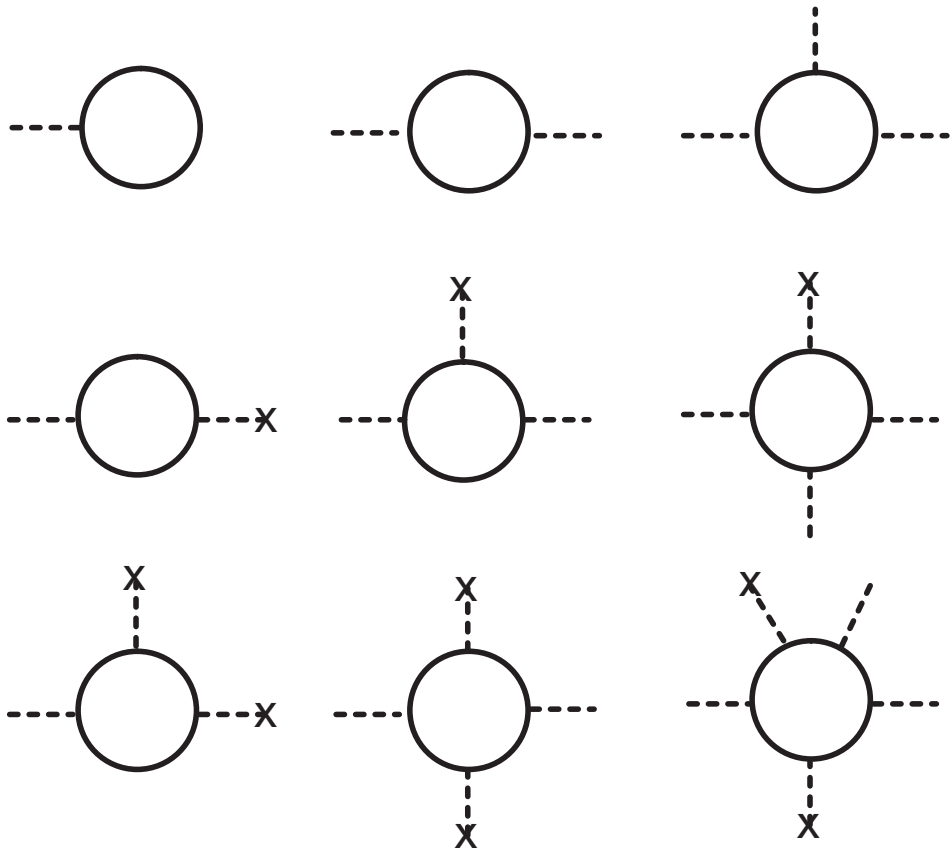}
   \caption{\small Diagrams which result when the f\mbox{}irst
   exponential operates on the functional integral (\ref{def1}).
   Full lines represent fermion propagators, dashed lines with a cross 
   $\sigma\to\mbox{vacuum}$ transitions, dashed lines stand for either  
   $\sigma$ or $\phi$ f\mbox{}ields.} 
\label{fig2}
\end{figure}

To integrate the fermions out of the Lagrangian ${\cal L}_1$ and not
break its transformation properties with respect to the chiral group, 
we modify the corresponding functional integral, writing it as
\begin{equation} 
\label{def1} 
   \int\!{\cal D}\bar{\psi}{\cal D}\psi\left\{\exp\left(
   \int\!\!\!\int\!\prod_{i=1}^{2} {\rm d}^2x_i  
   \frac{\delta}{\delta\sigma
   (x_1)}\Delta_\sigma (x_1-x_2)\frac{\delta}{\delta\Gamma (x_2)}
   \right) \exp\left(
   -i\!\int\! {\rm d}^2x[\bar{\psi}(i\!\!\not\!\partial +\widehat{m} 
   +\sigma + i\gamma_5\phi )\psi  
   +\frac{\widehat{m}}{g^2}\Gamma ]\right)\right\}_{|\Gamma =0}
\end{equation}
Obviously, this def\mbox{}inition coincides with the standard formula
at $\widehat{m}=0$. For $\widehat{m}\neq 0$ the action of the 
f\mbox{}irst exponent upon the second is to include as basic element 
in the functional integral the chiral symmetric tree-level result
shown in Fig. \ref{fig1}. On the other hand, by f\mbox{}irst
integrating out the fermions, this exponent generates an 
inf\mbox{}inite number of all possible one-loop fermion diagrams
attached to the external auxiliary f\mbox{}ields, including those that 
contain the $\sigma\to\mbox{vacuum}$ transitions, rendering the result 
invariant under chiral transformations. Some of these diagrams are
shown in Fig. \ref{fig2}. 

Indeed, taking into account eq. (\ref{prop}), one easily integrates 
over $x_2$ in (\ref{def1}). Then, using the formula 
\begin{equation}
   \left(\frac{\delta}{\delta\Gamma (x_1)}\right)^n
   \exp\left(-i\!\int\!{\rm d}^2x\,\frac{\widehat{m}}{g^2}\,\Gamma (x)
   \right) = \left(-i\frac{\widehat{m}}{g^2}\right)^n 
   \exp\left(-i\!\int\!{\rm d}^2x\,\frac{\widehat{m}}{g^2}\,\Gamma (x)
   \right),
\end{equation}
we derive for (\ref{def1}) an expression, where after setting $\Gamma =0$
and taking the f\mbox{}irst exponent out of the functional integral we get 
\begin{equation} 
\label{def2} 
   \exp\left(-\widehat{m}\!\int\!{\rm d}^2x\,
   \frac{\delta}{\delta\sigma (x)}\right)
   \int\!{\cal D}\bar{\psi}{\cal D}\psi  
   \exp\left( -i\!\int\! {\rm d}^2x[\bar{\psi}(i\!\!\not\!\partial 
   +\widehat{m} + \sigma + i\gamma_5\phi )\psi ]\right).
\end{equation}
Since the f\mbox{}irst exponent is a translation operator, which shifts 
the argument $\sigma\to \sigma
-\widehat{m}$, we obtain f\mbox{}inally for (\ref{def1})  
\begin{equation} 
\label{def3} 
   \int\! {\cal D}\bar{\psi} {\cal D}\psi  
   \exp\left( -i\!\int\! {\rm d}^2x[\bar{\psi}(i\!\!\not\!\partial 
   + \sigma + i\gamma_5\phi )\psi ]\right) \propto 
   \left(\det D_E\left|_{\widehat{m}=0}\right.\right)^N.
\end{equation}

Thus, we arrive at the following  bare ef\mbox{}fective action 
\begin{equation}
\label{Seff}
   \mbox{Re}\, S^{{\rm eff}}_E =\frac{N}{2}\mbox{Tr}\ln D^\dagger_E
   D_E |_{\widehat{m}=0} -\int\! {\rm d}^2x_E
   \frac{(\sigma -\widehat{m})^2+\phi^2}{2g^2}\, ,
\end{equation}
where the real part of the fermion determinant is invariant with
respect to the chiral transformations (\ref{dsigma})-(\ref{dphi}), as we 
wished to find, and 
$\delta {\cal L}_{{\rm eff}} =\widehat{m}\delta\sigma /g^2$, as it
follows from the second term.
 
One might think of criticizing the above calculations by claiming that
the same result (that is Eq. (\ref{Seff})) can be obtained by the
usual method, {\it i.e.,} by shifting $\sigma\to\sigma -\widehat{m}$
in (\ref{GN2}) and subsequently integrating over the 
fermions. We argue, however, that the shift of the scalar 
f\mbox{}ield in the generating functional, being a replacement of 
the variable, changes automatically the transformation law of the 
pseudoscalar $\phi$ to $\delta\phi =-2\theta (\sigma - \widehat{m})$. 
As a consequence, the symmetry breaking pattern of the
ef\mbox{}fective Lagrangian obtained in this way dif\mbox{}fers from 
(\ref{dL2}) and, therefore, suf\mbox{}fers from the problem indicated
in Sec. II. In contrast, Eq. (\ref{Seff}) obtained through the 
complementary tadpole diagrams contained in the def\mbox{}inition 
(\ref{def1}) leaves the auxiliary f\mbox{}ields and their
trasformation properties unchanged.


\section{Tadpoles at the level of the generating functional}

The procedure considered above can be easily implemented in the
framework of the functional integral approach. Indeed, writing the 
linearized Lagrangian density ${\cal L} (\psi ,\bar\psi ,\sigma
,\phi)$ as the sum of the symmetry breaking, ${\cal L}_{\rm SB}(\psi, 
\bar\psi )$, and symmetric, ${\cal L}_{\rm S}$, parts, one has for the 
generating functional
\begin{equation}
\label{Z1}
   Z = \!\int\! {\cal D}\psi {\cal D}\bar\psi \exp 
       i{\cal S} (\psi ,\bar\psi )
     = \!\int\! {\cal D}\psi {\cal D}\bar\psi             
       \, e^{i{\cal S}_{\rm SB}(\psi, \bar\psi )} \!\int\! 
       {\cal D}\sigma {\cal D}\phi \exp i{\cal S}_{\rm S} 
       (\psi ,\bar\psi,\sigma ,\phi),                
\end{equation}
where ${\cal S}_{\rm SB}(\psi, \bar\psi )=-\!\int\!{\rm d}^2x\,
\bar\psi\widehat{m}\psi$. At this stage, to carry out the $1/N$ 
expansion, one usually integrates over the fermions assuming that the 
order of integration in (\ref{Z1}) can be changed without serious
consequences. This is not true, actually. The reason is that the
double integral over bosonic f\mbox{}ields is a chiral invariant, but 
the one over fermions is not. Thus, the pattern of symmetry breaking 
will be altered. 

We argue now that the consistent procedure here should be as follows
\begin{equation} 
\label{Z2} 
   Z = \!\int\! {\cal D}\sigma {\cal D}\phi              
       \, e^{i{\cal S}_{\rm SB}(\sigma )} 
       \!\int\! {\cal D}\psi {\cal D}\bar\psi 
       \exp i{\cal S}_{\rm S} 
       (\psi ,\bar\psi ,\sigma ,\phi ),               
\end{equation}
{\it i.e.,} the change of the order of integrations in (\ref{Z1})
should be accompanied with the corresponding replacement of the
symmetry breaking term: ${\cal L}_{\rm SB}(\psi, \bar\psi )\to 
{\cal L}_{\rm SB}(\sigma ) =\widehat{m}\sigma /g^2$, where 
${\cal L}_{\rm SB} (\sigma )$ is the solution of the equation $\delta 
{\cal L}_{\rm SB} (\sigma )=\delta {\cal L}_{\rm SB} (\psi, \bar\psi
)$. This equation can be solved, because chiral transformation 
properties of bosonic and fermionic f\mbox{}ields are mutually 
correlated. As a result the functional integral over fermions in Eq. 
(\ref{Z2}) is chirally invariant (excluding the anomaly, which is not 
important for the question studied here), and therefore the symmetry
breaking pattern is strictly traced.   

At f\mbox{}irst sight, Eq. (\ref{Z2}) looks as if we did the shift 
$\sigma\to\sigma -\widehat{m}$ in (\ref{Z1}) to obtain it. This is not
the case. As we have already learned, the shift does not change the 
symmetry breaking pattern of the fermionic part of the Lagrangian, and 
therefore cannot lead to the desired chirally symmetric functional
integral over fermions. Eq. (\ref{Z2}) should instead be considered
either as a postulate related with the symmetry or as the result of
the $\sigma$-tadpole mechanism. The latter is in our understanding the
reason for the unconventional rule associated with the order of 
integrations in $Z$. This equation contains the main message of our
work because it expresses the result of Sect. \ref{sec.tm} in the 
shortest and most general form.

One can recognize in the replacement ${\cal L}_{\rm SB}(\psi, \bar\psi
)\to{\cal L}_{\rm SB}(\sigma )$ the use of the constraining relation 
(\ref{sfe}). On one hand, this explains why Eq. (\ref{Z2}) is inwardly 
consistent with (\ref{Z1}). On the other hand, if one accepts that it
is this f\mbox{}ield equation that controls the step from (\ref{Z1})
to (\ref{Z2}), the whole procedure can be straightforwardly applied to 
any theory with broken discrete $\gamma_5$-symmetry. The massive GN
model \cite{Zee:1997} is an example of such a theory. In this
particular case our formula leads to the result obtained by Feinberg 
and Zee for the ef\mbox{}fective potential.   
    
Next, let us try to understand in simple terms what is the main
dif\mbox{}ference between our calculations here and the method 
presented in \cite{Osipov:2000}. Following those works, we would
integrate out the fermions directly in Eq. (\ref{Z1}). Of course, we
would not pay much attention to the order of integrations there. 
However, we would calculate systematically (in the framework of the
heat kernel expansion) the counterterms which are needed for
consistency with the requirements of chiral symmetry
and would add them to the ef\mbox{}fective Lagrangian. What would be
the result? The answer is very simple. It would coincide with
the ef\mbox{}fective Lagrangian which one f\mbox{}inds integrating out 
the fermions in Eq. (\ref{Z2}). To see this it is instructive to look
again at Eq. (\ref{def1}), where the f\mbox{}irst exponent, being
taken out of the integral and applied to the result of integration, 
generates the inf\mbox{}inite set of counterterms which are necessary
to recover the correct symmetry breaking pattern of the outcome. It is 
exactly this what has been done in \cite{Osipov:2000}, but with the
use of a dif\mbox{}ferent and more complicated technique. It is clear 
that both results coincide.       


\section{Final comments and conclusion}

The renormalization of the ef\mbox{}fective Lagrangian (\ref{Seff})
can be done similarly to the massive GN model \cite{Zee:1997}. In
order that chiral symmetry constraints be valid in the renormalizable
theory, it is necessary that all divergences of the theory may be 
absorbed in two constants, the mass $\widehat{m}$ and the coupling 
constant $g^2$. It is also important that the boson variables are
chosen such that their chiral transformations do not depend on 
$\widehat{m}$, otherwise one would have fatal inf\mbox{}inities in the 
symmetry transformation laws. These conditions are fulf\mbox{}illed for
(\ref{Seff}) and we conclude that our calculations do not destroy the
renormalizability of the multi-fermion Lagrangian in two space-time 
dimensions.   
  
The method presented here can be easily extended to the known
$4$-dimensional models with nonabelian chiral symmetry and
multi-fermion interactions. Several examples are given by the 
Nambu -- Jona-Lasinio type models of QCD with explicit chiral
symmetry breaking. In these cases the original quark Lagrangian
satisfies the hypothesis of partially conserved axial vector current 
(PCAC) at the quark level and the tadpole mechanism which we suggest
protects the theory from loosing this property in the large $N_c$
limit. There are several examples in the literature where the subtle
but essential dif\mbox{}ference between Eqs. (\ref{Z1}) and (\ref{Z2}) 
was not recognized, leading to an unjustif\mbox{}ied violation of the 
PCAC relation. Note also that the lattice calculations for the massive 
GN model show that PCAC is satisf\mbox{}ied in the continuum limit 
\cite{Ichinose:1999}.   

To conclude, we have outlined the basic steps required to formulate
the functional integration over multi-fermion interactions with
explicit chiral symmetry breaking in leading $1/N$ order. Without loss
of generality we have considered the massive GN model, which is being 
actively studied nowadays \cite{Thies:1993}-\cite{Barducci:1995}. We 
have shown that the method preserves the symmetry breaking pattern of
the original Lagrangian and does not alter its dynamics. The information 
over all counterterms which are needed to be added to fulf\mbox{}ill
these requirements is cast into a very compact form and readily
implemented as a rule in the functional integration.  

\vspace{0.5cm}
{\bf Acknowledgements}
This work has been supported in part by grants provided by 
Funda\c c\~ao para a Ci\^encia e a Tecnologia, 
POCI/FP/63930/2005 and POCI/FP/81926/2007. This research is part of
the EU integrated infrastructure initiative Hadron Physics project 
under contract No.RII3-CT-2004-506078. 


\end{document}